\documentstyle[emulateapj,onecolfloat]{article}

\input epsf

\newcommand{\beq}{\begin{equation}}
\newcommand{\eeq}{\end{equation}}
\newcommand{\beqa}{\begin{eqnarray}}
\newcommand{\eeqa}{\end{eqnarray}}

\def\simlt{\lesssim}
\def\simgt{\gtrsim}

\newcommand{\ApJ}{ApJ}
\newcommand{\ApJL}{ApJ Lett.}
\newcommand{\ApJS}{ApJ Suppl.}
\newcommand{\PRL}{Phys. Rev. Lett.}
\newcommand{\PRD}{Phys. Rev. D}
\newcommand{\MNRAS}{MNRAS}

\newcommand{\AsAs}{A\&A}

\newcommand{\aut}[2]{{#1, #2.}}
\newcommand{\laut}[2]{{#1, #2.}}
\newcommand{\refs}[6]{#5, #2, #3,  {#4}.}

\newcommand{\mybib}[2]{\bibitem[#1]{#2}}

\begin{document}
\twocolumn[
\title{Sample Variance Considerations for Cluster Surveys}

\author{Wayne Hu and Andrey V. Kravtsov}

\affil{Center for Cosmological Physics and Department of Astronomy and
Astrophysics, University of Chicago, Chicago IL 60637}


\begin{abstract}
  We present a general statistical framework for describing the effect of
  sample variance in the number counts of virialized objects and examine its effect 
  on cosmological parameter estimation. Specifically, we consider effects of 
  sample variance on the power spectrum normalization and properties of dark energy extracted
  from current and future local and high-redshift samples of clusters.
  We show that for future surveys that probe ever lower cluster masses and
  temperatures, sample variance is generally comparable to or greater
  than shot noise and thus cannot be neglected in deriving
  precision cosmological constraints.  For example, sample variance is
  usually more important than shot variance in constraints on the
  equation of state of the dark energy from $z < 1$ clusters.
  Although we found that effects of sample variance on the
  $\sigma_8-\Omega_m$ constraints from the current flux and
  temperature limited X-ray surveys are not significant, they may be
  important for future studies utilizing the shape of the temperature
  function to break the $\sigma_8-\Omega_m$ degeneracy.  We also
  present numerical tests clarifying the definition of cluster mass
  employed in cosmological modelling and accurate fitting
  formula for the conversion between different definitions of halo
  mass (e.g., virial vs. fixed overdensity).

\end{abstract}

\keywords{}
]
\section{Introduction}

Clusters of galaxies are the most massive and rare virialized
objects in the universe. At the high mass end, their local abundance
depends exponentially on the {\it rms} of matter fluctuations and is
thus a very sensitive probe of the power spectrum normalization and
matter density in the universe (e.g., \cite{Evr89}\ 1989;
\cite{Freetal90}\ 1990; \cite{HenArn91}\ 1991; \cite{Lil92}\ 1992;
\cite{WhiEfsFre93}\ 1993; \cite{EkeColFre96}\ 1996; \cite{ViaLid96}\ 1996; 
\cite{Hen00}\ 2000; \cite{PieScoWhi01} 2001).  Large cluster
masses and high intracluster gas temperatures make galaxy clusters
observable in X-rays out to relatively high redshifts ($z\lesssim 1$).
Moreover, future Sunyaev-Zeldovich (SZ) surveys will provide
homogeneous cluster samples with relatively simple selection functions
out to $z\sim 3$ (e.g., \cite{Holetal00} 2000). The SZ surveys can
therefore be used to study evolution of cluster abundance over an
unprecedented range of redshifts. The abundance of clusters above a
certain mass in a given area of the sky as a function of redshift is
very sensitive to the amplitude and growth rate of perturbations as well
as the comoving volume per unit redshift and solid angle. These, in turn, are
sensitive to the cosmological parameters (e.g., \cite{HaiMohHol01}
2001). Therefore, strong constraints on the cluster normalization,
matter content, and the equation of state of the Universe can, in
principle, be obtained from large cluster surveys (\cite{HolHaiMoh01}\
2001; \cite{WelBatKne01} 2001).

The potential of future cluster surveys, however, can only be realized
after careful studies of all possible sources of theoretical and
observational errors and biases. First and foremost, it is critical to
understand the relation between observed and theoretical measures of
cluster mass The mass-observable relations depend on the processes
that shape the bulk properties of the intracluster medium (ICM).
Extensive simulations of cluster formation have shed light onto the
role of non-gravitational processes such as galactic feedback (e.g.,
\cite{MetEvr94}\ 1994; \cite{NavFreWhi95}\ 1995; \cite{BiaEvrMoh01}\ 2001; 
\cite{HolCal01}\ 2001) and radiative cooling (e.g.,
\cite{Muaetal01} 2001). Both feedback and cooling appear to bring the
predicted scaling relations between cluster X-ray luminosity,
temperature, and mass in closer agreement with observations. However,
both processes are notoriously difficult to incorporate into
simulations and further numerical effort and detailed comparisons with
new high-quality {\sl Chandra}, {\sl XMM-Newton}, and SZ observations
will be needed to test whether these processes alone can explain
various properties of the cluster ICM and its evolution.  As was
recently emphasized by \cite{Sel02} (2002) and \cite{Ikeetal01}
(2001), even the error budget on the local abundance is dominated by
the uncertainties in conversion from theoretical to observable
measures of mass.

Next in importance is the relationship between the cluster abundance
in mass and cosmology.  Recently, the use of very large volume
cosmological simulations has led to significant improvements in our
knowledge of the theoretical cluster mass function (\cite{SheTor99}\ 1999; 
\cite{Jenetal01}\ 2001; \cite{Evretal02}\ 2002; \cite{Zheetal02}\ 2002). 
In particular, \cite{Jenetal01} (2001) found that the mass
function at cluster scales can be described by a universal analytic
fit for all cosmologies and redshifts to an accuracy of $\lesssim
10-20\%$ in amplitude; this result was confirmed by \cite{Evretal02}
(2002) and \cite{Zheetal02} (2002).  While there is reasonable hope
that further studies will result in even more accurate
parameterizations of the mass function, even the current uncertainty
is not the principle remaining component of the error budget in
cosmological constraints.  \cite{Evretal02} (2002) found that it is
dominated by the uncertainties in conversion in mass definitions and
sample (cosmic) variance.  Indeed, the proliferation of different
cluster mass definitions in theoretical analyses may be a source of
considerable confusion and error in comparing predictions to
observations (\cite{Whi01} 2001).  Finally, unlike the other sources
of errors discussed above, sample variance is usually neglected in
theoretical and observational analyses.

In this paper, we present a statistical framework for describing the
sample variance for a cosmological population of virialized objects
identified at a single epoch or over a range of redshifts.  We apply
this formalism to evaluate the effects of the sample variance on the
current and future cosmological constraints, focusing on the specific
cases of constraints from the local cluster temperature function and
the evolution of the cluster abundance in future SZ surveys (e.g.,
\cite{Holetal00} 2000).  We show that uncertainties due to sample
variance are comparable to or greater than the Poisson errors in many
cases of future interest. Therefore, the variance should not be
neglected in analyses aiming to derive precision cosmological
constraints.  Although we found that effects of sample variance are
not significant in the interpretation of the local cluster abundance
from the current flux and temperature limited X-ray surveys to
determine the power spectrum normalization, it may be important for
future studies based on the shape of the temperature function
(\cite{Ikeetal01} 2001).  We also present discussion and numerical
tests clarifying the definition of mass in theoretical analyses and
provide useful fitting formulas for conversions between different mass
definitions.

The paper is organized as follows.  In \S \ref{sec:statistical} we
introduce the statistical formalism needed to describe sample variance
and its effect on cosmological parameter estimation.  We discuss
cluster scaling relations and survey selection in \S \ref{sec:survey}.
In \S \ref{sec:local} and \S \ref{sec:highz}, we discuss the impact of
sample variance on the interpretation of the local cluster abundance
and future high redshift surveys dedicated to studying the properties
of the dark energy.  Finally, in a series of Appendices, we give
details of survey window calculations, numerical tests of the cluster
mass function and bias and convenient fitting formulas for conversion
between different mass definitions.

\section{Statistical Formalism}
\label{sec:statistical}
\subsection{Sample Covariance}

Consider the general case of 
a population of objects selected for some property, for definiteness say 
their mass $M$ in some range $dM$ 
and assume that their number
density fluctuation is related to the underlying linear mass field by a
linear bias parameter
\begin{equation}
\delta n(M) = \bar n (M) b(M) \delta\,.
\end{equation}
The average number density within a set of windows
$W_i({\bf x}_i)$ becomes 
\begin{equation}
n_i = \int d^3 x_i W_i ({\bf x}_i) [ \bar n + \delta n_i ]\,,
\end{equation} 
where we assume that the windows are normalized so that 
$\int d^3 x W_i =1$.
The sample covariance of these estimates of the true number density
becomes
\begin{eqnarray}
{\langle n_i n_j \rangle - \bar n^2 \over \bar n^2} & = &
 b^2 \int d^3 x_i \int d^3 x_j W_i ({\bf x}_i) W_j({\bf x}_j) \nonumber\\
&& \quad \times \langle \delta({\bf x}_i) \delta({\bf x}_j) \rangle 
\\
&=& b^2 \int {d^3 k \over (2\pi)^3} W_i({\bf k}) W_j^*({\bf k}) P(k)\,.\nonumber
\label{eqn:variancesimp}
\end{eqnarray}
Note that the large-scale structure of the universe makes the number density
even in non-overlapping volumes covary. 
On  top  of this sample covariance one adds the usual shot noise variance which
goes as $1/\bar n V$ fractionally. 

This expression is easily generalized to the two cases of interest
where one wishes to extract cosmological information from the number density
as a function of mass and/or redshift.  In this general case, consider the 
covariance to be between windows at distinct redshifts and the selection to
be for distinct masses, $n_i(M_i,z_i)$ and $n_j(M_j,z_j)$.  Then
\begin{eqnarray}
{\langle n_i n_j \rangle - \bar n_i\bar n_j
\over \bar n_i \bar n_j} 
&=& 
b(M_i,z_i) b(M_j,z_j) D(z_i) D(z_j)\nonumber\label{eqn:variancegen}\\
&&\quad\times
\int {d^3 k \over (2\pi)^3} W_i({\bf k}) W_j^*({\bf k}) P(k)\,,
\end{eqnarray}
where $D$ is the linear growth rate and we have again assumed a deterministic
linear bias.

We will assume the following meaning of the terms {\it sample
  covariance, sample variance} and {\it cosmic variance}.
For a given sample of objects, we will call the covariance between its
subsamples (e.g.  Eqn.~[\ref{eqn:variancegen}] for subsamples of
clusters in different mass or redshift bins) sample covariance. We
will call the net effect of the covariance on a quantity of interest
(e.g. a cosmological parameter) sample variance.  Finally, for a
sample that encompasses the entire observable universe, sample
variance becomes cosmic variance.

\subsection{Survey Windows}
\label{sec:window}

The calculation of the sample covariance relies on the description of the
survey windows and the choice of binning in redshift and in mass.
A few simple cases are of interest for estimation purposes.  
Since in the consideration of windows, the distinction in 
Eqn.~(\ref{eqn:variancegen}) of redshift and mass bins does not enter,
we will consider the notationally simpler case of a single redshift and
mass range in Eqn.~(\ref{eqn:variancesimp}) without
loss of essential generality.

\begin{figure}[t]
\centerline{ \epsfxsize=3.2truein\epsffile{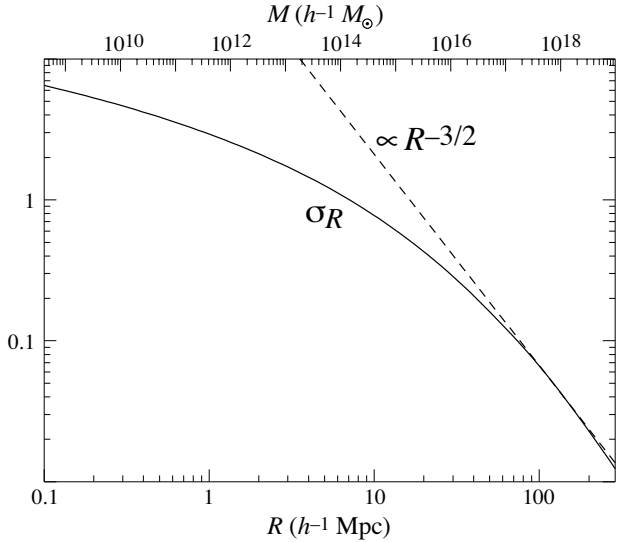} }
\caption{\footnotesize Root mean square fluctuation in the linear density field $\sigma_R$ in the
fiducial $\Lambda$CDM cosmology (\S \ref{sec:cosmology}) as
a function of top hat radius and the corresponding enclosed mass  
$M = (4\pi/3)\rho_{m} R^3$.  Because of the flatness of the linear power spectrum 
around $R=100 h^{-1}$Mpc, the rms scales in the same way as shot noise,
$V^{-1/2} \propto R^{-3/2}$ in this regime.  The mass $M_*$ 
at which $\sigma_R=1.69$ controls the relative strength of all effects
(see Figure~\ref{fig:variance}).}
\label{fig:sigma}
\end{figure}

For local surveys which cover a large fraction of the sky, it is often
a reasonable approximation to take the window as a single spherically
symmetric top hat volume of radius $R$.  Then the variance in
Eqn.~(\ref{eqn:variancesimp}) becomes
\begin{equation}
\sigma^2_n \equiv {\langle n^2 \rangle - \bar n^2\over \bar n^2}  = b^2 \sigma_R^2\,,
\end{equation}
where $\sigma_R$ is the familiar top-hat rms of linear density fluctuations
field.  This quantity is plotted for the fiducial $\Lambda$CDM cosmology
of \S \ref{sec:cosmology} in Fig.~\ref{fig:sigma}.  Also plotted is the
scaling of a white (or shot) noise power spectrum.  Because the linear
power spectrum is nearly flat for $R \approx 100 h^{-1}$ Mpc, sample variance
and shot variance scale with volume in a similar fashion for typical volumes. 
The sampling errors on this scale exceeds 10\% for a typical bias of
a few.

This basic example is readily generalized to the partially
overlapping multiple spherical windows of a flux or magnitude limited
selection and serves as a useful and simple order of magnitude 
estimate of sampling effects.

For surveys confined to a smaller section of sky but extending to
cosmological distances, the windows can be divided up into slices
in redshift.  Consider a series of slices in redshift at comoving 
distance $r_i$ and
width $\delta r_i$ with a field of radius $\Theta_s$ in radians
in the small angle approximation and a flat spatial geometry:
\begin{equation}
W_i({\bf k}) = 2 e^{i k_\parallel r_i} {\sin k_\parallel \delta r_i/2
        \over k_\parallel \delta r_i/2}  
        {J_1(k_\perp r_i \Theta_s) 
        \over k_\perp r_i \Theta_s}\,.
        \label{eqn:pillwindow}
\end{equation}
In the limit of slices that are thick compared with the coherence,
\begin{equation}
{\langle n_i n_j \rangle - \bar n^2 \over \bar n^2}  = \delta_{ij} b^2 
{1 \over \delta r_i r_i^2}
\int {d^2 l \over (2\pi)^2} 
\left(
        2{J_1(l \Theta_s) 
        \over l \Theta_s}\right)^2 P({l/r_i})\,.
\label{eqn:pillbox}
\end{equation}
Note that the angular window in the parenthesis goes to unity for
$l\Theta_s \ll 1$.

Eqn.~(\ref{eqn:pillbox}) is closely related to the Limber equation in
Fourier space (e.g., \cite{Kai92} 1992) in that there is no covariance
between distinct redshift bins and only modes perpendicular to the
line-of-sight enter into the variance.  We give a more general
consideration of the survey window including sky curvature and a
non-trivial angular mask and radial selection in the
Appendix~\ref{sec:gorywindow} (see Eqn.~[\ref{eq:goryvariance}]).

\subsection{Statistical Forecasts}
\label{sec:fisher}

While the calculation of the covariance in the previous sections can tell us
about the relative importance of sample and shot noise variance for various
choices of mass and redshift binning, it does not directly translate
into the relative importance on cosmological parameter estimation
where the weighting of the bins reflects parameter sensitivity.

\cite{HolHaiMoh01} (2001) introduced a useful Fisher matrix approximation to a 
Poisson likelihood analysis which essentially propagates the errors in the bins
to the covariance of more fundamental parameters.  Here we generalize the
technique to include sample covariance.  Define the Fisher matrix in the 
parameter space of interest $p_\alpha$ as
\begin{equation}
F_{\alpha\beta} = \sum_{ij} 
{\partial \bar n_i \over \partial p_\alpha} 
({\bf C}^{-1})_{ij}
{\partial \bar n_j \over \partial p_\beta} \,,
\end{equation}
where the covariance matrix includes both sample covariance and shot variance
\begin{equation}
C_{ij} = \langle n_i n_j \rangle - \bar n_i\bar n_j + \delta_{ij} \bar n_i/V_i\,.
\end{equation}
Then an estimate of the covariance matrix of the parameters follows as
\begin{equation}
C_{\alpha\beta} \approx ({\bf F}^{-1})_{\alpha\beta}\,.
\end{equation}
The Fisher approximation allows a rapid exploration of the parameter space(s)
and the effect of survey specifications on statistical errors.  
It has been shown to be a reasonable approximation 
for most cosmological parameters of interest
(\cite{HolHaiMoh01} 2001).  Note that the parameter set can include
``nuisance parameters'', e.g. an amplitude and a power law index for
an observable vs. mass relation.  Utilizing the Fisher matrix approximation,
these parameters would be marginalized at the expense of increasing
the errors
\begin{equation}
\sigma_\alpha = (C_{\alpha\alpha})^{1/2}
\end{equation}
on the parameters of interest.  Marginalization can be offset by prior 
knowledge in the form of an inverse covariance weighted sum of the
information
\begin{equation}
{\bf C}_{\rm tot} = ({\bf C}^{-1} + {\bf C}^{-1}_{\rm prior})^{-1},
\end{equation}
or loosely speaking by summing the Fisher matrices from individual
sources.  Because an infinitely sharp prior on all parameters save 
an $m$ element subset corresponds to considering the reduced
$m \times m$ Fisher sub-matrix, we term the corresponding errors
on the parameters the ``unmarginalized errors".  In the case of
$m=1$ parameter, the unmarginalized error is simply
$1/\sqrt{F_{\alpha\alpha}}$.

Finally, for the construction of priors and the interpretation of errors 
it is useful to note that under a re-para\-meter\-iza\-tion of the space
to the set $\pi_\mu(p_\alpha)$, the covariance matrix transforms as
\begin{equation}
C_{\mu\nu} = \sum_{\alpha\beta} {\partial \pi_\mu \over \partial p_\alpha}
                C_{\alpha\beta} {\partial \pi_\nu \over \partial p_\beta}\,.
\end{equation}
We will use this relation to evaluate the covariance matrix involving parameters
that depend on the fundamental set, e.g. $\sigma_8$ and the Hubble constant.

\section{Cluster Surveys}
\label{sec:survey}

We now specialize the sample variance considerations to cluster
surveys in the context of spatially flat CDM models with dark energy.
In \S \ref{sec:cosmology}, we define the cosmological framework and
the parameters of interest that underly it.  We provide further tests
of the critical ingredients: the cluster mass function, mass
definition, and bias prescription in Appendix \ref{sec:sims}.  In \S
\ref{sec:scaling}, we relate cluster observables to halo mass,
providing convenient conversion formulae between different definitions
of halo mass in Appendix \ref{sec:massconv}.  These relations are used
to define common selection criteria in \S \ref{sec:selection}.

\subsection{Cosmological Model}
\label{sec:cosmology}

\begin{figure}[t]
\centerline{ \epsfxsize=3.2truein\epsffile{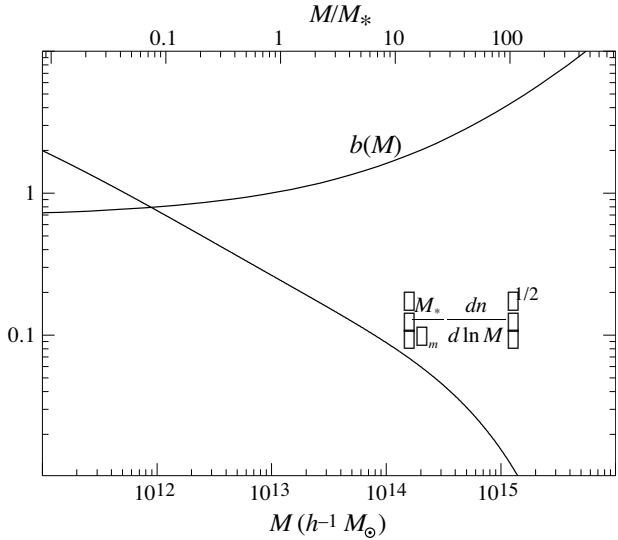} }
\caption{\footnotesize Mass function and bias in the fiducial $\Lambda$CDM cosmology
[see Eqns.~(\ref{eqn:massfun}) and (\ref{eqn:bias})] .
Although the bias increases for $M>M_*$ and enhances sample errors, the
exponential suppression of the number densities rapidly makes shot errors 
dominate at $M\gg M_*$.} 
\label{fig:bias}
\end{figure}

\begin{figure}[t]
\centerline{ \epsfxsize=3.5truein\epsffile{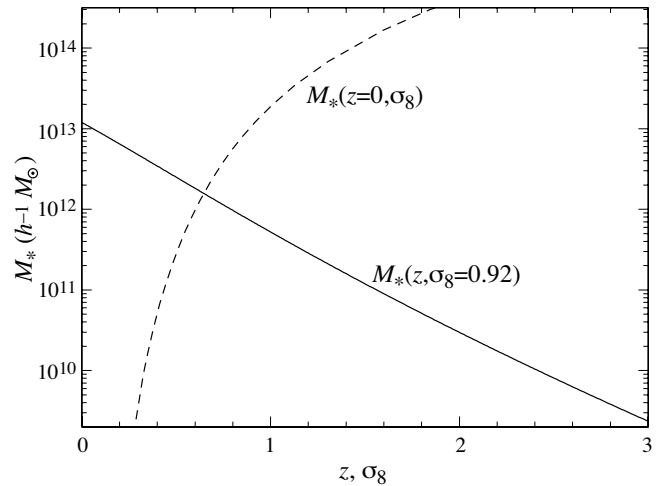} }
\caption{\footnotesize $M_{*}$ controls the relative importance of sample and shot variance  due to steeply declining number densities above $M_*$.  It is a
  strong function of redshift and an even stronger function of
  normalization.  The fiducial $\Lambda$CDM model assumes
  $\sigma_8=0.92$ at $z=0$ and hence $M_*(z=0)=1.2 \times 10^{13}
  h^{-1}$Mpc.  In conjunction with Eqn. (\protect\ref{eqn:scaling}),
  this figure can be used to determine the relative importance of
  sample variance for cases not explicitly explored here.}
\label{fig:mstar}
\end{figure}

We now specialize these considerations to cluster surveys.  We
associate galaxy clusters with dark matter halos of the same mass.
The mean differential comoving number density of dark matter halos
(\cite{Jenetal01} 2001, Eqn B3\footnote{We adopt the fit to the {\it
    unsmoothed} mass function of the halo catalogs with masses defined
  using the spherical overdensity of 180 with respect to mean density.
  This fit describes the actual abundance of clusters
  in simulations; smoothing artificially increases the amplitude of the
  mass function by $\sim 5-20\%$ (Jenkins et al 2001) and is
  unwarranted at the high cluster masses that we consider here.})
\begin{equation}
{d n \over d\ln M} = 0.301 {\rho_{m} \over M} {d \ln \sigma^{-1} \over d\ln M}
        \exp[-|\ln \sigma^{-1} + 0.64|^{3.82}]\,.
\label{eqn:massfun}
\end{equation}
and the linear bias (\cite{MoWhi96}\ 1997; \cite{SheTor99}\ 1999)
\begin{equation}
b(M) = 1 + {a \delta_c^2/\sigma^2 -1 \over \delta_c} 
         + { 2 p \over \delta_c [ 1 + (a \delta_c^2/\sigma^2)^p]}
\label{eqn:bias}
\end{equation}
are modeled from fits to cosmological simulations with $a=0.75$ and $p
= 0.3$ (the values are modified to match the mass function for
clusters in the Hubble volume simulations, R. Sheth, private
communication).  Here $\delta_c=1.69$ is the threshold overdensity of
spherical collapse in a matter dominated universe; we will ignore the
weak scaling of $\delta_c$ with cosmology.  $\sigma^2(M,z)$ is the
variance in the density field smoothed with a top hat that encloses a
mass $M$ at the mean matter density today (or equivalently the
comoving density) $\rho_{m}$.  The correspondence in the adopted mass
function and bias is to the mass enclosed in a region where the
density is 180 times the mean matter density, as demonstrated in
Appendix \ref{sec:sims}.  The main cosmological scaling of both the
number density and bias comes through $\sigma$.  It is useful then to
scale the mass to $\sigma(M_*) = \delta_c$.  $M_*(z)$ is shown in Fig.
\ref{fig:mstar}.

We consider the fiducial cosmology to be spatially flat and
defined by 6 independent parameters:
the dark energy density $\Omega_{\rm DE}=0.65$, equation of state
$w=-1$, physical matter density $\Omega_m h^2= 0.148$,
physical baryon density $\Omega_b h^2 = 0.02$, tilt $n_s=1$,
and the initial normalization of the 
curvature fluctuations $\delta_\zeta = 4.79 \times 10^{-5}$ at $k=0.01$ 
Mpc$^{-1}$ to match the COBE detection.
The non-standard definition of the normalization $\delta_\zeta$ 
is given explicitly in
\cite{Hu01c} (2001) and removes the cosmological parameter dependence
of the more traditional choices.  For the fiducial model, it converts
to $\delta_H= 4.42 \times 10^{-5}$, $\sigma_8=0.92$,
or $M_* = 1.2 \times 10^{13} h^{-1} M_\odot$. 
We shall see that all results are a strong function of the 
assumed normalization but can be rescaled through $M_*$
(see Fig. \ref{fig:mstar} and Eqn.~(\ref{eqn:scaling})).

\subsection{Scaling Relations}

\label{sec:scaling}

The mass of a cluster is usually not measured directly, but 
estimated from models of its X-ray luminosity,
emission weighted temperature $T_X$, velocity dispersion,
or, in the future, its Sunyaev-Zel'dovich 
integrated flux.  

Following \cite{Ikeetal01} (2001), we will assume an $L_X-T_X$ relation of
\begin{equation}
L_X = 1.23 \times 10^{37} \left({T_X \over 6 {\rm keV}} \right)^{2.5} 
        h^{-2}\, {\rm W m}^{-2} \,.
\label{eqn:LxTx}
\end{equation}
Here $L_X$ is the X-ray luminosity in the energy range of
$[0.1-2.4]$~keV and $T_X$ is emission-weighted temperature of the
intracluster gas.  

The connection between theoretical predictions and observations 
is provided by the virial mass - temperature relation
$M_v-T_X$, which we parameterize as  
\begin{eqnarray}
\left( {M_v \over 10^{15} h^{-1} M_\odot} \right) &= &
\left[ {\beta \over (1+z)(\Omega_m\Delta_v)^{1/3}}{T_X \over 1 {\rm keV}} \right]^{3/2}, 
\end{eqnarray}
where $\Delta_v$ is 
the virial overdensity with respect to the mean matter density
(see Eqn. \ref{eqn:deltav}). 
$\beta$ parameterizes the scaling and in general may be a function of
$T_X$ and cosmology.

Numerical simulations by various groups give $\beta \approx 0.75$
(e.g. \cite{EvrMetNav96}\ 1996; \cite{EkeNavFre98}\ 1998;
\cite{BryNor98}\ 1998; also \cite{PieScoWhi01}\ 2001 for a tabulation of
results), roughly independent of temperature and cosmology.
Observations favor a lower normalization $\beta \approx 0.54(T_X/6{\rm
  keV})^{0.24}$ (\cite{FinReiBoh01}\ 2001)
making the observed cluster sample less massive.  
Here we have converted from
the mass at a much higher overdensity of 500 times the critical 
density $M_{500/\Omega_m}$ to $M_v$.
The observed relation
consequently indicates a lower normalization $\sigma_8$ in a fixed cosmology
(\cite{Sel02}\ 2002).  We will employ the theoretical $M_v-T_X$ relation
to be consistent with our choice of cosmology; we will discuss effects
of different $M_v-T_X$ relation in \S\ref{sec:discussion}.
Since the Jenkins mass function is defined at $\Delta=180$, not at
$\Delta_v$, the virial $M_v-T_X$ relation must be further converted to
an $M_{180}-T_X$ relation to match with the mass function.
These conversions can be done using simple
but accurate fitting formulae for the general
interconversion of mass definitions assuming an NFW (\cite{NavFreWhi97}
1997) profile given in Appendix \ref{sec:massconv}.
  
The relationship between $M_v$ and our $M$ can be approximately 
parameterized as 
\begin{equation}
\left( {M \over 10^{15} h^{-1} M_\odot} \right) = 
(1+b_1)\left( {M_v \over 10^{15} h^{-1} M_\odot} \right)^{1+b_2}
\end{equation}
for $14 < \log(M/h^{-1} M_\odot) < 16$ and halo concentrations 
of the fiducial model with $b_1 = 0.168 x + 0.373 x^4$ and
$b_2 = 0.0113 x + 0.0176 x^4$ and $x \equiv 1-\Omega_m(z)$.
Note that in the matter-dominated limit the two mass definitions
coincide.

For the fiducial $\Omega_m=0.35$ cosmology the $M-T_X$ relation becomes
\begin{equation}
\left( {M \over 10^{15} h^{-1} M_\odot} \right) = 
\left[ {1.113 \beta \over (1+z)(\Omega_m\Delta_v)^{1/3}}{T_X \over 1 {\rm keV}} \right]^{
1.516} \,.
\label{eqn:MTx}
\end{equation}
Because this $10\%$ correction in temperature will propagate into a
$50\%$ correction of a flux-limited survey volume at a fixed mass,
this seemingly small correction is required.  Note however that the
difference between simulation results for $\beta$ is at least
comparable to this shift and the discrepancy between the theoretical
fits to $M-T_X$ in simulations and observed relation
(\cite{HjoOukvan98}\ 1998; \cite{HorMusSch99}\ 1999; \cite{NevMarFor00}\ 2000; 
\cite{FinReiBoh01}\ 2001; \cite{AllSchFab01}\ 2001) is more
substantial still.

\subsection{Survey Selection}
\label{sec:selection}

The survey selection is a critical element in determining the
importance of sample variance.  Even ignoring errors in the conversion
of observables to mass, which always decreases the relative importance
of sample variance, the underlying cluster scaling relationships enter
into the consideration of the survey selection and hence the survey
window.  A survey may be limited by X-ray flux, SZ flux, and/or
temperature as well as volume or redshift and we will consider these
as the relevant selection criteria.

For the present purpose of determining
the ultimate limitation sample variance places on cluster surveys, we take
the scaling relations (\ref{eqn:LxTx}) and (\ref{eqn:MTx}) with $\beta=0.75$
to be both correct and deterministic.  In reality 
one should include in the final error budget the intrinsic scatter, 
measurement, and modeling uncertainties in the observables and their
relation to mass.    

The $M-T_X$ relation allows us to convert a temperature threshold, typically 
$T_X \simgt 1-3$keV into a mass threshold $M>M_{\rm th}$.  
The combined $M-T_X$ and $L_X-T_X$ relations convert mass to X-ray luminosity.
For sample with a flux limit $f_{\rm lim}$, the effective radius of the volume probed
by the survey is
\begin{equation}
R ={ d_L \over 1+z} = {1 \over 1+z}\sqrt{ L_X(M) \over 4\pi f_{\rm lim}}\,,
\label{eqn:rfluxlim}
\end{equation}
which implies that the survey window in mass bins will be overlapping 
concentric spheres or sky-wedges that increase in radius with the mass.  
The sample
variance calculation must then properly account for the covariance
between the mass bins.   
Note that the survey volume $V(M) \propto \beta^{-15/4}$ and hence strongly
depends on the $M-T_X$ relation adopted.
This flux limit is typically placed on top of a 
volume (redshift) limit that defines, for example, a local sample of clusters.

For SZ cluster surveys a flux decrement limit corresponds to $d_A^{-2}
T_{SZ} M \approx$ const for a sufficiently large angular diameter
distance $d_A$ such that the cluster is within the effective field of
view (\cite{HolCar01} 2001).  Under the assumption that the gas
density weighted electron temperature $T_{SZ} \approx T_X$, and the
$M-T_X$ relation above, an SZ flux limit corresponds to a mass
threshold that is roughly constant with redshift and scales as $M_{\rm
  th} \propto \beta^{3/5}$ for different $M-T_X$ relations.  Note that
clusters of a given mass are detectable to all relevant redshifts
despite the flux limit and details of the $M-T_X$ and $T_X-T_{SZ}$
relations are much less important than in a flux limited X-ray survey.
Nonetheless, the $M-T_X$ relation has a severe effect on survey yields
if the underlying power spectrum is normalized to an incorrect value
by a misinterpretation of the local X-ray cluster abundance (see \S
\ref{sec:discussion}).

\section{Local Abundance}
\label{sec:local}
We consider here the effect of sample variance compared to shot noise and
cosmological parameter degeneracy on the interpretation of the local cluster 
abundance in the classic $\sigma_8-\Omega_m$ plane (\cite{Evr89}\ 1989; 
\cite{Freetal90}\ 1990; \cite{HenArn91}\ 1991; \cite{Lil92}\ 1992; 
\cite{WhiEfsFre93}\ 1993).  
We begin with the idealized case of a volume and temperature
limited survey and then consider the more realistic X-ray 
flux, volume and temperature
limited case.

\subsection{Volume Limited Survey}
\label{sec:localvol}

Let us start with a toy model to develop intuition for the effects of
sample variance.  Consider a volume limited local survey with a radius
$R$ that detects all clusters above a limiting mass $M_{\rm th}$
determined by a temperature limit and the $M-T_X$ relation.

First consider binning the clusters into a single mass bin $M > M_{\rm th}$.   
There is then only one window to consider and we can directly compare the 
sample and shot noise errors on number density determinations. 
Note that sample variance which is equal to shot variance leads
to a $\sqrt{2} \approx 1.4$ increase in the errors.  

\begin{figure}[t]
\centerline{ \epsfxsize=3.5truein\epsffile{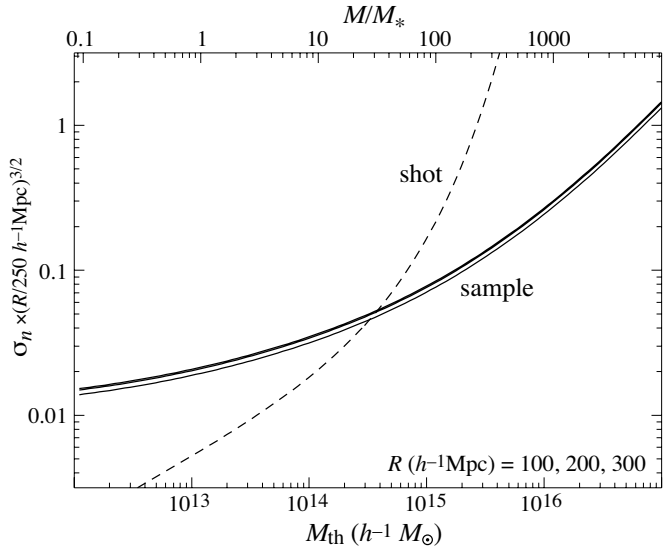} }
\caption{\footnotesize Sample vs. shot errors in number counts 
as a function of threshold mass for a local survey.   Sampling errors
dominate at $M < 30 M_* \approx 4 \times 10^{14} h^{-1} M_\odot$.
Errors have been scaled by $(R/250 h^{-1}$Mpc$)^{3/2}$ to reflect 
the local volume 
and the coincidence of the curves shows that both scale $V^{-1/2}$.  
Approximate results for alternate cosmologies or redshifts can be estimated by
the scaling with $M_*$ (see Fig. \protect\ref{fig:mstar} and Eqn.~(\ref{eqn:scaling})).}
\label{fig:variance}
\end{figure}

In Fig.~\ref{fig:variance}, we show the sample and shot errors for
$R=100, 200, 300 \, h^{-1}$Mpc.  As noted in \S \ref{sec:window}, in
the $R=100 h^{-1}$ Mpc range both errors scale as $V^{-1/2}$ and we
have removed this dependence by multiplying by $(R/250 h^{-1}$Mpc)$^{3/2}$.
The corresponding fiducial volume is $6.5 \times 10^7$ $h^{-3}$ Mpc$^{3}$,
comparable to the volume of actual local surveys.  Notice that the two
errors cross at $M_{\rm th} \approx 4 \times 10^{14} h^{-1}\,M_\odot$
or $M_{\rm th}/M_* \approx 30$ with sample variance dominating at the
lower masses.  This crossing mainly reflects the exponential suppression of 
the number densities for $M_{\rm th}/M_* \gg 1$ and hence the dramatic increase
in shot noise.  For small changes in 
redshifts, normalizations, 
and cosmologies these qualitative statement remain true but
require rescaling. Roughly,
\begin{equation}
{M_{\rm th} \over M_*} \simlt 31 
\left( { M_* \over M_{* \rm fid}}\right)^{-1/2} \left( \Omega_m \over 0.35 \right)^{1/2}
\left( 40 \sigma_{200} \over \sigma_8 \right)\, ,
\label{eqn:scaling}
\end{equation}
for sample to dominate shot variance, where recall 
$M_{* \rm fid} = 1.2 \times 10^{13} h^{-1} M_\odot$.  
For a low $\Gamma=\Omega_m h$ cosmology, the enhancement of sample 
variance from the excess large scale power 
(at typical scales of $R=200 h^{-1}$ Mpc)
from the shape of the power spectrum 
dominates and can substantially increase the importance of sample variance.
In conjunction
with Fig. \ref{fig:mstar}, this scaling may be used as a rough check
to determine whether sample variance is relevant to a given problem.
For surveys that reach this limiting $M_{\rm th}$, 
sample variance will dominate over shot noise
in the total number density  and lead to
a substantial increase in the statistical errors.

\begin{figure}[t]
\centerline{ \epsfxsize=3.5truein\epsffile{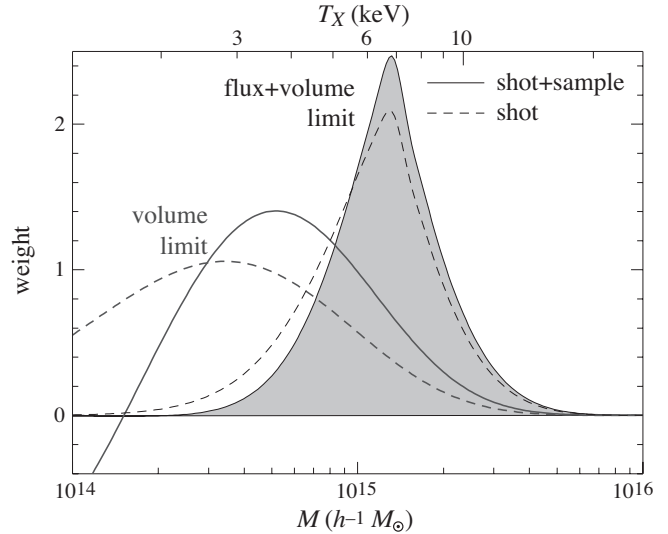} }
\caption{\footnotesize Mass range sensitivity of the normalization determination for surveys with
  a volume limit of $z_{\rm max}$ and mass threshold of $M=10^{14}
  h^{-1} M_\odot$ with and without an additional X-ray flux limit of
  $f_{\rm lim} = 2 \times 10^{-14}$ W m$^{2}$.  With shot noise only,
  the normalization constraint gets its most significant contribution
  for $3-4 \times 10^{14} h^{-1} M_\odot$ ($\sim 3$ keV) for the
  volume limit and $10^{15} h^{-1} M_\odot$ ($\sim 6-7$ keV) for the
  additional flux limit.  Correspondingly sampling errors are
  substantially more important for the former than the latter, also
  altering the optimal mass weighting (solid vs. dashed lines).}
\label{fig:weights}
\end{figure}

Now consider a survey that can also bin in masses, for example from individual
temperature measurements.   The relative contribution of sample variance will
depend on what range of masses is most relevant to the question at hand.  
This sensitivity in turn can be a function of survey selection and cosmology.
Here the Fisher matrix technique of \S  \ref{sec:fisher} is useful since it  automatically
folds in these factors in a minimum variance weighting.

As a simple example, consider the estimation of a 
single parameter  $p_\alpha$ 
with all others fixed by prior knowledge.   The minimum variance weighting 
of the $i$ mass bins is given by
\begin{equation}
w_i = {1 \over F_{\alpha\alpha}}
\sum_{j} {\partial \bar n_i \over \partial p_\alpha} 
(C^{-1})_{ij} {\partial \bar n_j \over \partial p_\alpha} \,.
\end{equation}
For a diagonal covariance matrix as in the case of shot noise, 
this simply reduces to an inverse variance weighting of mass bins.

For illustrative purposes, let us take 
a volume limit of $z=0.09$ ($R=270 h^{-1}\,$Mpc) and an all
sky survey.  We bin the masses from $\log (M/ h^{-1} M_\odot) = 14$ to $16$ in 40
steps of $0.05$.
The predicted mass function is sufficiently smooth that this binning more
than suffices to recover all the information contained in its shape.
Consider the parameter of interest to be the power spectrum normalization
$\delta_\zeta$ or $\sigma_8$.
In Fig.~\ref{fig:weights}, we show the weights assuming shot noise only
(dashed lines) and shot plus sample covariance (solid lines).  
The mass sensitivity peaks at about $3-4 \times 10^{14} h^{-1} M_\odot$ for
shot noise only
implying that the inclusion of sampling errors would produce
a substantial change in the error budget. 
Indeed the Fisher approximation 
yields a factor of 2 degradation in the errors on $\sigma_8$ or 
sample variance that is 3 times more important than shot variance.
Other parameters, when also considered individually (unmarginalized) 
show similar error degradations.
Note that the optimal weighting in the presence of sample covariance shifts
to higher masses and can have negative contributions due to the covariance
of the bins so that neglect of sample covariance in the analysis can degrade
errors even further.

\begin{figure}[t]
\centerline{ \epsfxsize=3.5truein\epsffile{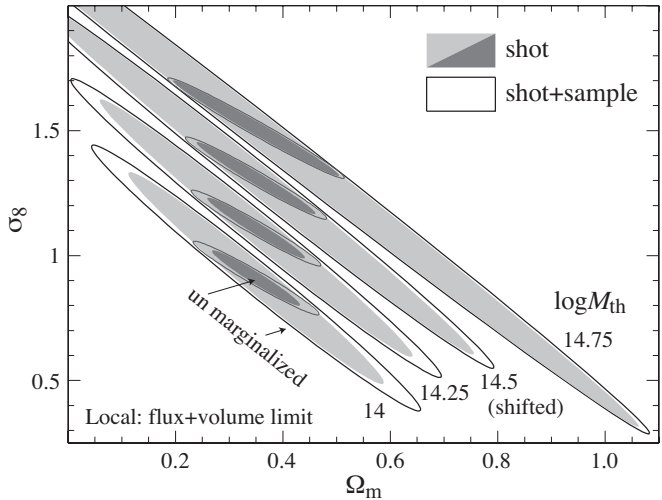} }
\caption{$\sigma_8-\Omega_m$ constraints from a local X-ray flux and volume limited survey 
  at $65\%$ CL ($z_{\rm max}=0.09$, $f_{\rm lim} = 2 \times 10^{-14}$
  W m$^{2}$).  Unshaded ellipses denote full shot and sampling errors
  while shaded ellipses denote shot errors only.  Outer ellipses show
  marginalization over other parameters (except $w$) with very weak
  priors, inner ellipses show the unmarginalized constraints.  The
  four sets of $M_{\rm th}$ correspond to $T_X=1.25$,$1.83$,$2.67$,
  $3.9$ keV with the fiducial $M-T_X$ relation.  The three higher
  choices of $M_{\rm th}$ have ellipses shifted in steps of $0.2$ in
  the $y$-direction for clarity.  The effect of sample variance
  decreases with increasing parameter degeneracies and mass
  threshold.}
\label{fig:sigom}
\end{figure}

\subsection{Flux Limited Sample}

More realistically, the sample of local clusters will be X-ray flux as
well as volume and temperature limited so that low mass clusters are
found only nearby.  In this case, the mass sensitivity shifts to
higher masses and sample variance becomes less important.

For illustrative purposes, let us take a flux limit of $f_{\rm lim} =
2 \times 10^{-14}$ W m$^{-2}$ in addition to the volume limit of
$z=0.09$ and mass threshold of $M_{\rm th}=10^{14} h^{-1} M_\odot$ 
($T_X =1.25$~keV),
parameters typical for the current flux-limited samples of clusters
with estimated X-ray temperatures (e.g., \cite{Ikeetal01}\ 2001).  The
mass weighting for the normalization parameter now peaks around
$10^{15} h^{-1} M_\odot$ or temperatures around $T_X = 6-7$keV with
only a small difference if sample variance is included with shot
variance.  The degradation in normalization errors is consequently
reduced to 20\%.  Different parameters have different sensitivities.
For the chosen parameter set (see \S\ref{sec:cosmology}), the relative
degradation in the flux limited case is largest for $\Omega_{\rm DE}$
where it reaches 40\%.  Thus even for a flux limited survey, sample
variance can be equal to shot variance in the statistical errors.

The overall importance of sample variance can decrease due to other
sources of errors in the estimation, including degeneracies in the
parameters and errors in mass estimation, especially at 
low masses or temperatures.  To illustrate these effects, consider 
the usual two dimensional parameter space $\sigma_8-\Omega_m$ by
transforming our fundamental space as described in \S \ref{sec:fisher}.
In Fig. \ref{fig:sigom}, we show the effect of sample variance 
on the 68\% confidence region in this plane for several choices of
the minimum mass (or temperature) 
in $\log(M_{\rm th}/ h^{-1} M_\odot)$ of the clusters employed, offset for the higher
minimum masses for clarity.  Larger ellipses represent marginalization over
other parameters (with very weak priors of $\sigma_{\Omega_b h^2}=0.01$ and
$\sigma_{n_s}=0.3$ mainly to eliminate unphysical degeneracies)
and smaller ellipses represent unmarginalized constraints 
or prior-fixing of the other parameters.  Both sets have a prior of 
$\sigma_w = 0$
since the dark energy equation of 
state is highly degenerate with $\Omega_m$.

\begin{figure}[t]
\centerline{ \epsfxsize=3.5truein\epsffile{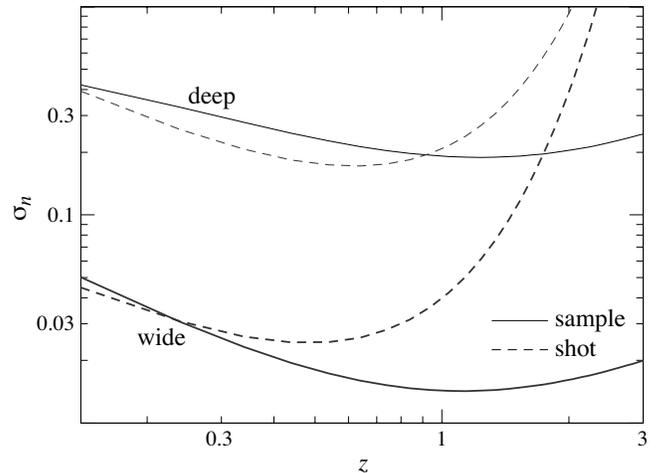} }
\caption{\footnotesize Shot and sampling errors on the number density $n$ as a function of redshift
  for the deep ($M_{\rm th} = 10^{14} h^{-1}\, M_\odot$, 12 deg$^{2}$)
  and wide configurations ($M_{\rm th} = 2.5 \times 10^{14} h^{-1}\,
  M_\odot$, 4000 deg$^{2}$). For the deep survey, sampling errors
  dominate shot errors out to $z\sim 0.9$, whereas for the wide
  survey, they cross at $z \sim 0.2-0.3$.  The importance of sample
  variance to cosmological constraints depends on their sensitivity in
  redshift relative to these crossover points.  Approximate results for
  other cases of interest can be scaled from Fig. \ref{fig:mstar} and
  Eqn.~(\ref{eqn:scaling}).}
\label{fig:errorz}
\end{figure}
Not surprisingly, sample variance becomes irrelevant as the mass
threshold is increased.  Sample variance also decreases in importance
when going between the fixed and marginalized cases due to the
dominance of parameter degeneracies in the latter.  For the $M_{\rm
  th}=10^{14}h^{-1} M_\odot$ threshold, the increase in the errors on
$\Omega_m$ go from $40\%$ in the fixed case to $30\%$ in the
marginalized case.  The effect of marginalization mainly comes through
parameters that change the shape of the linear power spectrum.
Realistic current constraints on these parameters places reality
approximately half-way in between the two extremes shown in
Figure~\ref{fig:sigom}.  Finally, note that the errors in the
direction orthogonal to the degeneracy line increase negligibly with
the addition of sample variance and threshold mass $M_{\rm th}$.
Since the local cluster abundance is usually used to constrain this
direction (c.f. \cite{Ikeetal01}\ 2001 for use of the degenerate
direction), its cosmological interpretation is essentially unmodified
by sample variance or uncertainties at the low mass end.

\begin{figure*}[t]
\centerline{ 
        \epsfxsize=6.5truein\epsffile{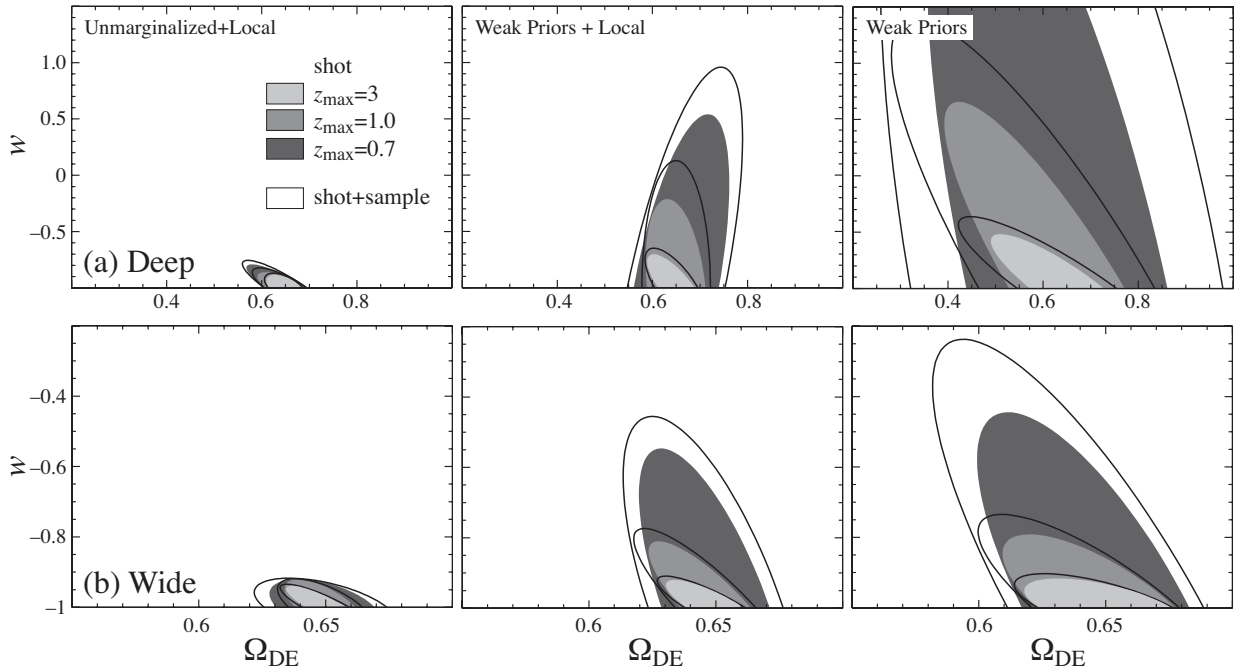}
}
\caption{\footnotesize Dark energy constraints (68\% CL) for the (a) deep and (b) wide high redshift
  cluster surveys.  Panels right to left reflect the addition of
  priors from the base set of weak conservative priors to the addition
  of a local cluster abundance constraint to the further addition of
  sharp priors on all other parameters.  Three different maximum
  redshifts are shown with and without the inclusion of sample
  variance errors.}
\label{fig:deep}
\end{figure*}

\section{High-$z$ Surveys}
\label{sec:highz}

The number abundance of high redshift clusters is widely recognized as
being extraordinarily sensitive to cosmological parameters due to the
exponential cut off in the mass function above $M_*(z)$
(\cite{BahFan98}\ 1998; \cite{BlaBar98}\ 1998; \cite{ViaLid99}\ 1999).
Recently, its use in constraining the density and equation of state of
the dark energy using planned Sunyaev-Zel'dovich (SZ) cluster surveys has
been the focus of several studies (\cite{HaiMohHol01}\ 2001; 
\cite{WelBatKne01}\ 2001).  These studies show that if the shot noise
of rare high redshift clusters were the only source of uncertainty,
the SZ effect can constrain the equation of state at the percent
level.  Here we consider the effects of additional uncertainties from
sample variance and parameter degeneracies 
for two classes of planned surveys: a deep but narrow
survey volume and a wide but shallower survey.

\subsection{Fiducial Surveys}

As discussed in \S \ref{sec:selection}, an SZ survey has the virtue of
being able to detect clusters of a sufficiently high mass out to all
redshifts that they exist.  In addition, the survey selection can be
described by an effective threshold mass $M_{\rm th}$ that depends
only weakly on redshift (\cite{Holetal00} 2000).  Following
\cite{HolHaiMoh01} (2001), we take as our fiducial surveys a 12
deg$^{2}$ survey with a step function mass selection at $M_{\rm th}=
10^{14} h^{-1}$ Mpc (``deep") and a 4000 deg$^{2}$ survey to $M_{\rm
  th}=2.5 \times 10^{14} h^{-1}$ Mpc (``wide").  We take a minimum
redshift of $z=0.09$ to match the maximum redshift of the local
surveys of the previous section and bin in redshift to $\Delta z=0.1$.
The evolution of the number abundance is sufficiently smooth for the
constant $w$ models considered that finer binning is not needed, as we
find no significant degradation of constraints when varying bin size
up to $\Delta z=0.1$. This is comfortably larger than the expected
accuracy of individual cluster redshifts estimated using photometric
redshifts of member galaxies (e.g. \cite{Feretal02} 2002).  We calculate the sample covariance
under the window approximation of Eqn.~(\ref{eqn:pillwindow}) in both
cases.  Since
the redshift binning corresponds to scales of $\sim 3000 \Delta z
h^{-1}$ Mpc, the covariance is confined to a few neighboring redshift bins
but still should not be neglected.  Likewise, the further Limber-like 
approximation of Eqn.~(\ref{eqn:pillbox}) is not valid for the wide survey
since line-of-sight modes contribute comparably to transverse modes
for these large transverse scales. 
In addition, sky curvature begins to play a role for the wide survey
and a more detailed analysis should employ the general window expression
(Eqn.~[\ref{eq:goryvariance}]) given in Appendix~\ref{sec:gorywindow}.
We neglect this correction since we are mainly interested in the
relative effects of sampling errors.

In Fig.~\ref{fig:errorz} we show the relative contribution of shot and
sample errors as a function of the redshift of the bin.  Because of
the low mass threshold of the deep survey, sampling errors dominate
all the way out to $z \approx 0.9$.  For the wide survey, the crossing
occurs at $z \approx 0.2-0.3$ but sample variance remains important at
$z \simlt 0.7$.  
The crossing point for other cases of interest may be scaled from
Eqn.~(\ref{eqn:scaling}) and Fig.~\ref{fig:mstar}.
Whether sample variance affects cosmological
constraints depends on where in redshift the parameter sensitivity is
maximized.

\begin{figure*}[t]
\centerline{ \epsfxsize=6.75truein\epsffile{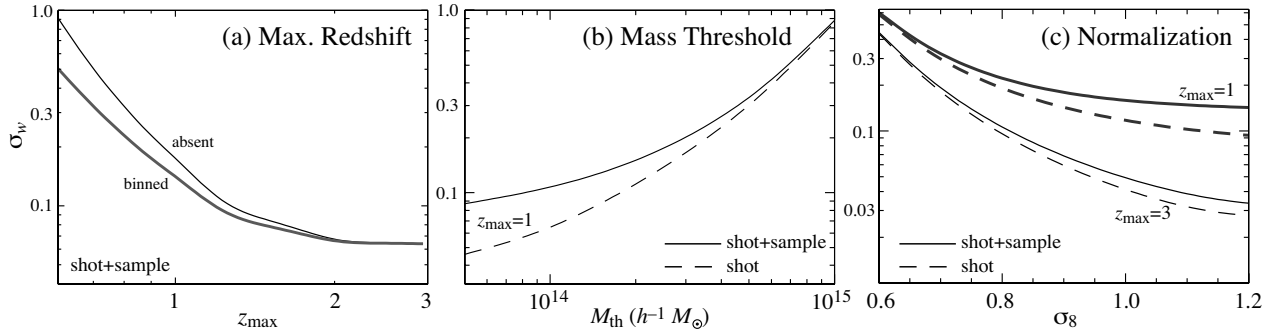} }
\caption{\footnotesize Exploration of sensitivity of equation of state errors to
  assumptions in the wide survey and cosmology.  (a) Maximum redshift
  with $z>z_{\rm max}$ either absent in the survey catalog or binned
  into a single mass bin. (b) Mass threshold for a $z_{\rm max}=1$ and
  (c) normalization of the linear power spectrum $\sigma_8$ for
  $z_{\rm max}=1$ and $3$.  Weak priors are assumed throughout.}
\label{fig:sigw}
\end{figure*}

Even though SZ surveys have no redshift limit, the selection may still
be redshift limited for two reasons: redshift followup and
uncertainties in evolution.  Since the SZ effect is redshift
independent it also gives no intrinsic handle on the redshift of
clusters so discovered.  Fortunately, the relative crudeness of the
binning required ($\Delta z \approx 0.1$) implies that photometric
redshifts of cluster member galaxies should be more than sufficient.
Still, photometric redshifts in the range $1 \simlt z \simlt 2.5$ are
difficult to obtain from optical ground-based observations.  We shall see that
even if redshifts cannot be determined beyond a maximum redshift
$z_{\rm max}$ much of the cosmological information is recovered simply
by binning all such clusters into a single high redshift bin.
More worrying is any evolution in cluster 
properties that change $M_{\rm th}$ (or more generally the mass selection 
function)
which cannot be accurately modeled from simulations or detailed 
multi-wavelength observations.  In this case the clusters above a certain
$z_{\rm max}$ may be unusable for cosmological constraints.

\subsection{Dark Energy Constraints}

Let us now specialize these considerations to the question of
constraining the density $(\Omega_{\rm DE}$) and equation of state ($w$)
of the dark energy.  Constraints in this plane and the relative
importance of sample variance are far more sensitive to prior
assumptions on other cosmological parameters than the normalization
from the local abundance.  We choose three levels of prior knowledge:
(1) Weak priors that reflect a conservative view of current
cosmological constraints $\sigma_h = 0.065$, $\sigma_{\Omega_b
  h^2}=0.01$, $\sigma_{n_s} =0.3$, $\sigma_{\ln \zeta} = 2$, (2) Weak
priors plus a local cluster abundance constraint from \S
\ref{sec:local} for the flux and volume limited case, and (3)
Unmarginalized errors with the addition of the local cluster abundance
constraint.

In Fig.~\ref{fig:deep}, we show the resulting constraints for the deep
and wide surveys and maximum redshifts of $z_{\rm max}=0.7$, $1.0$ and
$3$.  For the deep surveys and weak priors, degradations in the errors
on ($\Omega_{\rm DE}$, $w$) change slightly from (1.3,1.3) to (1.4,1.4) to
(1.4,1.3) as $z_{\rm max}$ increases.  The degradation decreases
[(1.3,1.3) to (1.1,1.4) to (1.1,1.2)] once the local abundance is
added to a high $z_{\rm max}$ make the constraints less dependent on
the sample variance limited low redshift end.  Fixing other parameters
reduces the effect still further.  Note that the effect of priors on
the absolute errors is much larger than on the relative effect of
sample variance.  For the wide survey the degradation decreases from
(1.3,1.4) to (1.1,1.2) as $z_{\rm max}$ increases with weak priors and
changes to (1.2,1.2) to (1.2,1.1) with the addition of the local
abundance constraint.  These smaller degradation factors reflects the
overall decrease in the importance of sample variance with increasing
mass threshold.  In summary, sample variance causes a 10-40\%
degradation in dark energy constraints depending on the detailed
assumption made.  Since 40\% reflects equal shot and sample variance
contributions, sample variance cannot be ignored in the statistical
error budget of future surveys.  Moreover it can affect the optimal
survey strategy given fixed observing constraints, e.g. making it more
important to study rare high redshift clusters.

Let us further explore these issues in the context of the wide survey,
weak priors and the measurement of the equation of state parameter
$w$.  Both the absolute constraints on $w$ and the relative importance
of sample variance are very sensitive to three quantities: the maximum
redshift $z_{\rm max}$, the mass threshold $M_{\rm th}$, and the
fiducial value of the normalization $\sigma_8$ (see
Fig.~\ref{fig:sigw}).  Note the sharp increase in the errors for
$z_{\rm max} < 1$.  If the maximum redshift simply reflects a lack of
followup redshift information beyond this redshift, a fair fraction of
the lost information can be regained just by binning all such clusters
into a single high redshift bin.  A low mass threshold at high
redshift is also critical for obtaining strong constraints on $w$ and
also increases the relative importance of sample variance.  Finally
the assumed normalization of the cosmological model has a strong
effect on the errors on $w$ due to the lack of high redshift clusters
as the normalization decreases.  A lowering of the normalization by a
factor of $0.8$ changes the errors on $w$ by nearly a factor of 3.
Lowering the normalization also reduces the importance of sample
variance as clusters become rarer, especially at high redshift.

\section{Discussion and Conclusions}
\label{sec:discussion}

The analyses presented in the preceding sections showed that for
  future surveys that probe ever lower cluster masses and
  temperatures, sample variance is generally comparable to or greater
  then the shot noise and thus cannot be neglected in deriving
  precision cosmological constraints.  For example, sample variance
typically doubles the statistical uncertainty in constraints on the dark
energy equation of state in a SZ survey limited to $z_{\rm max} \simlt
1$.  Its relative importance is very weakly dependent on survey volume
and is mainly a decreasing function of mass threshold of the survey,
$M_{\rm th}$, compared with the mass of a typical halo, $M_*$, because
of the rarity of high mass clusters.  We emphasize though that it is
simply that the shot noise variance for future surveys is impressively
small, and the cosmological prospects correspondingly bright, that the
small absolute effect of sample variance plays any role at all!

Although we found that effects of sample variance on the
$\sigma_8-\Omega_m$ constraints from the current flux and temperature
limited X-ray surveys of local clusters are not significant, they may
be important for future studies utilizing shape of the temperature
function to break the $\sigma_8-\Omega_m$ degeneracy, where sample
variance typically increases the statistical variance by a factor of
two.

We have studied the effects of sample variance under the assumption
that the fiducial $\Lambda$CDM cosmology is essentially correct and,
most importantly,  its joint normalization to the {\sl COBE} results
and cluster abundance assuming the usual simulation-based conversion
of X-ray temperature to mass.  Recent improvements in the measurement
of the observed mass-temperature relation has called the latter
assumption into question (\cite{FinReiBoh01}\ 2001, \cite{Ikeetal01}\ 2001).  
Because of the extraordinary sensitivity of
cluster survey yields to the normalization of the power spectrum,
these developments can significantly alter the prospects of cluster
surveys including the relative importance of sampling variance.
Employing the observed relation from \cite{FinReiBoh01} (2001) and
fixing the number of clusters at $T_X>6$ keV in our idealized flux and
volume limited surveys, we find that the normalization of the fiducial
cosmology would be lowered to $\sigma_8=0.75$ from $0.92$, in
agreement with conclusions of \cite{Sel02} (2002).

This change has opposite consequences for the importance of sample
variance for the local and high-redshift surveys.  For the local
surveys, since the number density in temperature is fixed, the
decrease in the mass threshold, $M_{\rm th}$
increases the importance of sample variance (see
Eqn.~[\ref{eqn:scaling}]). For example, degradation in the
unmarginalized errors on $\Omega_{\rm DE}$ increases from $1.4$ to $1.6$.
Sample variance effects would be even larger if the model is also
tilted to match the {\sl COBE} normalization.  For the high-redshift
surveys, the weak scaling of mass threshold with the mass-temperature
relation (see \S~\ref{sec:selection}) and strong scaling of
constraints with normalization (see Fig.~\ref{fig:sigw}) make sample
variance less important compared to our analyses.  More importantly,
the change in the normalization relation would substantially degrade
the ability of SZ surveys to measure dark energy properties due to the
increase in the rarity of high redshift clusters.  These relative
changes\footnote{Sample variance becomes more important in absolute
  terms at high $\Omega_{\rm DE}$ but constraints on dark energy
  properties also improve.}  remain true if the mass-temperature
change indicates a higher dark energy density $\Omega_{\rm DE} \approx
0.75$, $\Omega_m h^2 = 0.13$, with a $\sigma_8=0.85$ that then matches
the COBE normalization as well as the $T_X>6$ keV cluster abundance.

This current ambiguity in predictions reflects the general point that
until the mass-observable relationships for clusters are well
understood with new observations and better simulations, the predicted
cosmological yield of future cluster surveys and the statistical error
budgets considered here should be taken as highly uncertain. These
relationships therefore should be the primary focus of the future
modeling efforts and can be attacked in several ways.  First, the use
of higher resolution and more sophisticated cluster simulations
together with new high-quality {\sl Chandra} and {\sl XMM-Newton}
observations should fuel progress in our understanding of cluster
scaling relations in the near future.  The search for causes of
discrepancy between the observed and predicted $M-T_X$ relations that
currently compromises interpretation of the local cluster abundances,
is already the subject of substantial observational and modeling
effort. At the same time, independent constraints on mass-observable
relations can, in principle, be obtained from weak lensing analyses of
large X-ray selected cluster samples. In addition, power spectrum
normalization at cluster scales can be obtained independently from
studies of cosmic shear (e.g. \cite{Waeetal02} 2002). Independent
measurement of $\sigma_8$ would allow use of local cluster temperature
function to constrain present-day $M-T_X$ relation. Although at higher
redshifts prospects are less promising for the immediate future,
deep X-ray and SZ follow-up observations to the SZ surveys should provide a
wealth of data on such cluster populations. These
data will be critical for comparisons with numerical simulations,
tests of the scaling relations evolution, and, ultimately, precision
cosmological constraints from cluster surveys.

{\it Acknowledgments:} We thank J. Carlstrom, Z. Haiman, G. Holder,
J.J. Mohr, C. Pryke, A. Evrard, and M. White for stimulating
conversations.  WH was supported by NASA NAG5-10840 and the DOE OJI
program.

\appendix

\section{Window Function}
\label{sec:gorywindow}

We outline here the calculation in the case of a general survey window.
Consider a general radial selection $R_i(r)$ and angular mask $\Theta(\theta,\phi)$
such that the set of windows satisfies the separability condition
$W_i({\bf x})\equiv R_i(r) \Theta(\theta,\phi)$.  The Fourier transform
of the window then becomes  
\begin{equation}
W_i({\bf k}) = 4\pi  \sum_{lm} (-i)^l  
                \tilde R_{i\, l}(k)\tilde \Theta_{lm} Y_l^m(\theta_k,\phi_k) 
\end{equation}
and ($\theta_k$,$\phi_k$) define the direction of ${\bf k}$ with respect
to a fiducial direction, e.g. the center of the window.
Here the spherical harmonic transform of the angular mask is
\begin{equation}
\tilde \Theta_{lm} = \int d\Omega Y_l^{m*}(\theta,\phi) \Theta(\theta,\phi)\,,
\end{equation}
and the spherical Bessel transform of the radial window is
\begin{equation}
\tilde R_{i\,l}(k) = \int r^2 dr j_l(kr) R_i(r)\,.
\end{equation}
The covariance becomes
\begin{equation}
{\langle n_i n_j \rangle - \bar n^2 \over \bar n^2}  =  b^2
\sum_{lm} 4\pi \int {d k\over k} \tilde R_{i\,l}(k) \tilde R_{j\,l}(k)
 |\tilde \Theta_{lm}|^2 {k^3 P(k) \over 2\pi^2} \,.
\label{eq:goryvariance}
\end{equation}

In the Limber approximation of a slowly varying $P(k)$, one can
use the identity
\begin{equation} 
\int k^2 dk j_l(kr) j_l(kr') = {\pi \over 2r^2} \delta(r-r')\,,
\end{equation}
to show that the covariance is negligible
\begin{equation}
{\langle n_i n_j \rangle - \bar n^2 \over \bar n^2}  =  \delta_{ij} b^2
\int r^2 dr R_i^2(r) 
\sum_{lm} 
|\tilde \Theta_{lm}|^2 P(l/r) \,.
\end{equation}
Further specializing to azimuthally symmetric windows with
a normalized radial tophat profile,
\begin{equation}
{\langle n_i n_j \rangle - \bar n^2 \over \bar n^2}  =  
\delta_{ij}b^2 {1 \over r_i^2 \delta r_i}  
\sum_l {2 l+1 \over 4\pi} 
\tilde \Theta_l^2 P(l/r)\,,
\end{equation}
where the angular window is
\begin{equation}
\tilde \Theta_l =
2\pi \int_{-1}^{1} d \cos \theta P_l(\cos\theta) 
                \Theta(\theta)\,.
\end{equation}
In the limit that $l\gg 1$ this expression agrees with the
pillbox window Eqn.~(\ref{eqn:pillbox}). Note that the angular
windows in both expressions 
are normalized to unity for $l \Theta_s \ll 1$, where $\Theta_s$
is the typical angular dimension of the window.

\section{Numerical tests}
\label{sec:sims}

\begin{figure}[t]
\centerline{ \epsfxsize=3.5truein\epsffile{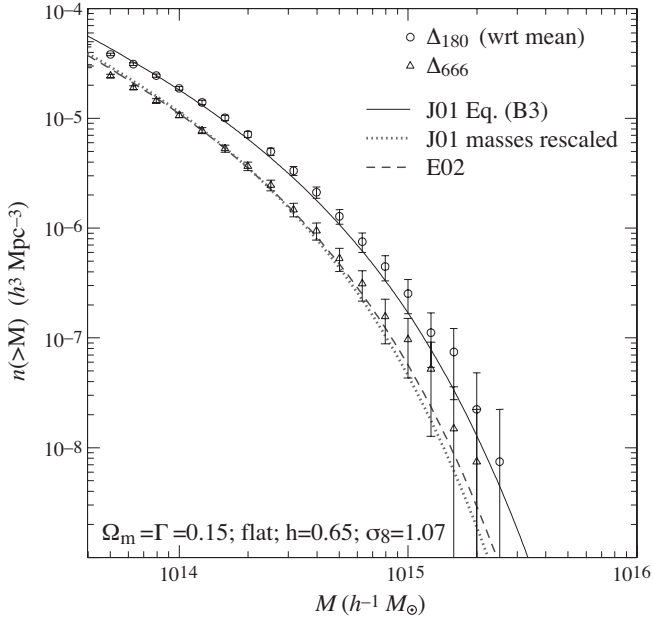} }
\caption{\footnotesize Cluster mass functions derived from the
  simulation of a flat low-$\Omega_m$ CDM cosmology
  ($\Omega_m=1-\Omega_{\rm DE}=\Gamma=0.15$; $h=0.65$; $n_s=1$;
  $\sigma_8=1.07$) using masses defined by $\Delta_{180}$ and
  $\Delta=666.6$. The error bars show $2\sigma$ Poisson errors.
  The {\em solid line} shows the fit obtained by \cite{Jenetal01}
  (2001, their Eqn. B3) while the dotted line shows this fit with
  $M_{180}$ converted to $M_{666}$, as described in the
  Appendix~\ref{sec:massconv}.  Agreement indicates that the \cite{Jenetal01}
  fit should be interpreted as $M_{180}$, defined with respect
  to the mean density and that the mass function is universal for this
  definition of mass.  The converted mass function fit matches both
  our simulated mass function for $\Delta=666$ and the fit of
  \cite{Evretal02}\ (2002, their Tab. 1) to the mass function from the
  Hubble volume $\Lambda$CDM simulation with this definition of mass.
  The agreement shows that conversion assuming an NFW profile
  adequately accounts for the substantial differences between
  $M_{180}$ and $M_{666}$.}
\label{fig:mfsim}
\end{figure}

During the past several years there has been significant advances in
our understanding of  halo mass function and bias from simulations
(e.g., \cite{MoWhi96} 1996; \cite{SheTor99} 1999; \cite{Jenetal01}
2001).  Nevertheless, there remains a certain confusion about the
meaning of mass in the mass function expressions, sufficiently
significant to make analyses of cluster abundances ambiguous
(\cite{Whi01} 2001).  Furthermore, bias models were extensively tested against
numerical simulations, but only for relatively small halo masses,
$M\lesssim 10-20 M_{\ast}$ (\cite{MoJinWhi97} 1997; \cite{Jin99} 1999;
\cite{SheTor99} 1999), which for our fiducial model corresponds to masses
$\lesssim 2.6\times 10^{14}h^{-1}\ {\rm M_{\odot}}$.  In cluster studies,
especially those at high redshift, halo masses $M>20 M_*$  must be 
interpreted and hence we test the bias model over the whole range.

In this Appendix we use results of a large $N$-body simulation to test
the mass function and halo bias expressions used in our analyses. The
simulation followed evolution of $256^3\approx 1.6\times 10^7$
particles in the flat CDM model with vacuum energy:
$\Omega_m=1-\Omega_{\rm DE}=0.15$; $h=0.65$; $n_s=1$; $\sigma_8=1.068$;
$\Gamma = 0.15$; $L_{\rm box}=512h^{-1}\ M_{\odot}$.  Here $\Gamma$ is the shape parameter in the adopted
approximation of the initial power spectrum (\cite{BBKS} 1986). The
simulation was run with the Adaptive Refinement Tree code
(\cite{KraKlyKho97} 1997) using $256^3$ zeroth level grid and five
levels of refinement.  The power spectrum normalization, $\sigma_8$,
was chosen to satisfy constraints from the local cluster abundance
(e.g., \cite{PieScoWhi01} 2001).

The adopted cosmological model is sufficiently different from the
models studied by \cite{Jenetal01} (2001) to make our analysis a
useful test of the universality of the mass function fit
advocated by these authors. In addition, differences
between various common definitions of halo mass increase for lower
values of $\Omega_m$. 
For example, consider the halo mass defined
within the radius corresponding to the {\it fixed\/} overdensity with
respect to mean density of the Universe,
$M_{180}=(4\pi/3)R_{180}^3 \Delta_{180} \rho_m$ where
$\Delta_{180}=180$ independent of cosmology.  It is 
is equivalent to the definition of mass with
respect to the cosmology-specific {\it virial overdensity\/} given by
the spherical collapse model, 
$M_v =(4\pi/3)R_v \Delta_{v}\rho_m$
only for $\Omega_m=1$ 
because then $\Delta_{180} =
\Delta_v$
(see Eqn.~\ref{eqn:deltav}).  For low $\Omega_m$, $\Delta_{180}$ and $\Delta_v$
are significantly different. For $\Omega_m=0.15$, for example,
$\Delta_v \approx 532$.
 This translates into $\approx 30\%$ difference in mass
for $M_{\rm vir}=5\times 10^{14}h^{-1}\ {\rm M_{\odot}}$ (see
Appendix~\ref{sec:massconv}).   This cosmology provides a sharp
test of the \cite{Jenetal01} (2001) assertion that the mass function 
of halos defined at $\Delta_{180}$ is independent of cosmology.

Figure~\ref{fig:mfsim} shows the halo mass functions derived from the
simulation using masses defined for $\Delta_{180}$ 
and $\Delta = 666.6$.  The latter definition was used by
\cite{Evretal02} (2002) to get fits to the mass functions in the
Hubble volume simulations (it corresponds to their definition of mass
at the overdensity of 200 with respect to the 
{\it critical density} for $\Omega_m=0.3$: $200/0.3\approx 666.6$). The solid line shows the fit obtained by
\cite{Jenetal01} (2001, their Eqn. B3 for this definition of
mass\footnote{Eqn. B3 gives fit to the unsmoothed mass function which
  is what we plot in Figure~\ref{fig:mfsim}.}), while the dotted line
shows this fit with $M_{180}$ converted to $M_{666}$, as described in
the Appendix~\ref{sec:massconv}. 
The figure clearly shows that halo mass in the
 universal Jenkins et al. fit should be interpreted as $M_{180}$,
 where the overdensity of 180 is defined with respect to the {\it mean density.}  
The converted mass function fit matches both our
simulated mass function for this overdensity and the fit of \cite{Evretal02}
 (their Table 1). While it is clear that the differences in mass
definition lead to substantial differences in the amplitude and shape
of the mass function, Fig.~\ref{fig:mfsim} shows that these
differences can be taken into account by the appropriate mass
conversion.

\begin{figure}[t]
\centerline{ \epsfxsize=3.5truein\epsffile{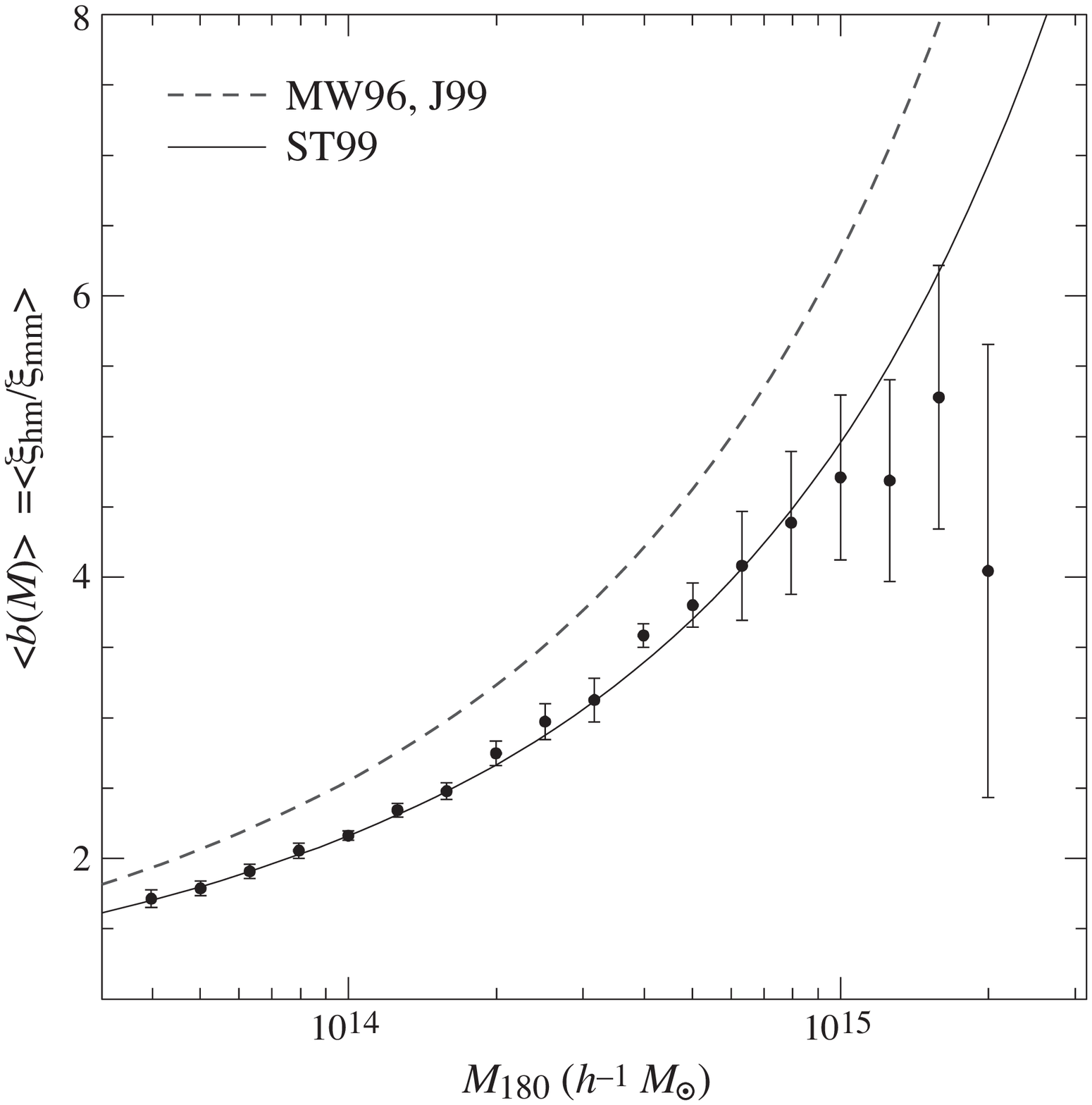} }
\caption{\footnotesize The average bias for halos of mass
  $>M_{180}$ predicted by the analytic models of \cite{MoWhi96} (1996,
  {\it dashed line}) and \cite{SheTor99} (1999, {\it solid line}), as
  averaged over the \cite{Jenetal01} (2001) mass function, and in the
  simulation (see Fig.~\ref{fig:mfsim}), estimated from the average
  ratio of halo-mass and mass-mass correlation functions at linear
  scales: $\langle b(>M)\rangle = \langle\xi_{hm}/\xi_{mm}\rangle_{\rm
    lin}$, as described in the text.  The $2\sigma$ error bars are the
  Poisson errors in $\xi_{mm}$ and $\xi_{hm}$ propagated to $b$. Note
  the points are not independent because they are derived from the
  same simulation and use overlapping cluster catalogs. }
\label{fig:bmsim}
\end{figure}

Figure~\ref{fig:bmsim} compares the average bias for halos of mass
$>M$ in simulations to the predictions of analytic models of
\cite{MoWhi96} (1996) and \cite{SheTor99} (1999). The bias in these
models in linear regime is given by Eqn~(\ref{eqn:bias}) with
$(p,a)=(0,1)$ and $(0.3,0.75)$, respectively. The analytic predictions
were computed as a mass function weighted average of the linear bias
predictions, with the mass function fit of Jenkins et al. (2001).
In the simulation, the bias was estimated as the average ratio of
halo-mass and mass-mass correlation functions at linear scales:
$\langle b(>M)\rangle = \langle\xi_{hm}/\xi_{mm}\rangle_{\rm lin}$.
Namely, we estimated $\xi_{mm}(r)$ and $\xi_{hm}(r, >M)$ for halos
with masses greater than $M$ (varying $M$ over the entire probed range
of masses; $M$ here is defined for $\Delta=180$). The bias was
estimated by averaging $\xi_{hm}/\xi_{mm}$ over $20-30h^{-1}\ {\rm
  Mpc}$ where bias is scale-independent. This gives us an estimate of
the mass function weighted bias as a function of halo mass.  The
figure shows that both models reproduce the mass dependence of halo
bias for cluster masses.  The value of the bias, however, is much
better matched by the model of \cite{SheTor99} (1999) in all but the
highest mass bins. The number of clusters in these bins is rather
small (the last mass bin contains only 3 clusters) and the deviations
are not significant.  The figure shows that Eqn.~(\ref{eqn:bias})
provides an accurate description of bias for the cluster mass halos.

\section{Mass Conversion}
\label{sec:massconv}

Various definitions of the mass of a halo can be converted assuming
a halo density profile.  Here we provide a fit to the general scaling
function that converts definitions under the assumption of
a NFW profile (\cite{NavFreWhi97} 1997).

The NFW halo profile has the functional form
\begin{equation}
\rho(r) = { \rho_s \over (r/r_s) (1+r/r_s)^2 }\,,
\end{equation}
for which the mass enclosed at a radius $r_h$ is
\begin{equation}
M_h = 4\pi \rho_s r_h^3 f(r_s/r_h)\,,
\end{equation}
where
\begin{equation}
f(x) = x^3 [\ln(1+x^{-1}) - (1+x)^{-1}]\,.
\label{eqn:massscaling}
\end{equation}
The limiting behavior of this function is
\begin{eqnarray}
\lim_{x\rightarrow 0} f(x) & = &  -[1+\ln(x)]x^3 \,,\nonumber\\
\lim_{x\rightarrow\infty} f(x) & =& {1 \over 2}x  - {2 \over 3}\,,
\end{eqnarray}
which says that if $r_h \gg r_s$ then up to a logarithmic
correction, the mass converges to that enclosed near the
scale radius $r_s$ and if $r_h \ll r_s$ the mass increases as $r_h^2$

A common baseline definition of mass for comparison is the
virial mass
\begin{eqnarray}
M_v  &\equiv& {4\pi r_v^3 \over 3}\Delta_v \rho_m \,,\nonumber\\
     &=& 4\pi \rho_s r_v^3 f(1/c) \,,
\end{eqnarray}
where the concentration parameter $c\equiv r_v/r_s$ and $\Delta_v$ is 
virial overdensity with respect to the {\it mean} matter density 
(\cite{BryNor98} 1998)
\begin{equation}
\Delta_v \approx { 18 \pi^2 + 82 x -39 x^2 \over  1+ x}\,,
\label{eqn:deltav}
\end{equation}
where $x \equiv \Omega_m(z)-1$.
We will give the conversion from 
an arbitrary definition of the halo mass to the virial mass although
the procedure is completely general. 

Defining the halo mass as 
\begin{equation}
M_h  \equiv {4\pi r_h^3 \over 3}\Delta_h \rho_m \,,
\end{equation}
where $\Delta_h$ can depend on cosmology but not on $M_h$, 
we relate the two radii as 
\begin{equation}
f(r_s/r_h) = {\Delta_h \over \Delta_v} f(1/c)\,,
\end{equation}
so that
\begin{eqnarray}
{r_s \over r_h}& = &x\left(f_h={\Delta_h \over \Delta_v} f(1/c) \right) \,.
\end{eqnarray}
Converting between definitions of mass simply involves inverting
$f(x)$ of Eqn.~(\ref{eqn:massscaling}) 
to find $x(f)$ and hence an explicit formula for the
relationship between two different definitions of $r_h$ or $M_h$
\begin{eqnarray}
{M_h \over M_v}& = & {\Delta_h \over \Delta_v} \left(r_h \over c r_s\right)^3\,.
\label{eqn:massmap}
\end{eqnarray}

The function $f(x)$ is monotonic its single argument $x$ and it is simple
to form a numerical lookup table for its inverse. Nonetheless we here
provide an accurate fitting formula for the inversion 
\begin{eqnarray}
x(f) &=& \left[ a_1 f^{2 p} + \left( {3 \over 4}\right)^2 \right]^{-1/2} + 2f \,,
\label{eqn:fit}
\end{eqnarray}
where $p = a_2 + a_3 \ln f + a_4 (\ln f)^2$ and the 4 fitting
parameters are 
$(a_1,\ldots,a_4) = (0.5116, -0.4283,-3.13\times 10^{-3}, -3.52\times 10^{-5})$.
This fit converts masses to better than $1\%$ accuracy across concentrations
$0 < c < 400$ ($2.5 < M_v/ h^{-1} M_{\odot} < \infty$ in the fiducial model;
for galaxy and cluster scales $c<20$, the conversion has typical errors of $\sim 0.3$\%).  
Note that the inversion is exact for $c \rightarrow 0$
($f \rightarrow \infty$) by construction.  

An alternate form that
is exact in the limit $c \rightarrow \infty$ can be obtained via
Taylor expansion as 
\begin{eqnarray}
{M_h \over M_v} = 1+ g(c,(\Delta_h/\Delta_v)^{1/3})\,,
\end{eqnarray}
where
\begin{eqnarray}
g(x,y) &=& 
 - 3(x+y)[x - x y + (1+x)(x + y) \ln (1+x) 
\nonumber\\ &&
- (1+x)(x+y)\ln(1+x/y)]
\nonumber\\&&
\times [3(1+x)(x+y)^2 \ln(1+x) \nonumber\\
&& - x(x+4x^2 + 6 x y + 3 y^2)]^{-1}\,,
\end{eqnarray} 
and for order unity $y$ exceeds the accuracy of
Eqn.~(\ref{eqn:fit}) for $c>10$.

These formulas allow for a convenient mapping of $M_v \rightarrow M_h$ 
once
$c(M_v)$ is specified.  Following \cite{Buletal01} (2001), we take 
\begin{equation}
c(M_v) = 9 (1+z)^{-1} (M_v/M_*)^{-0.13}
\end{equation}
where $M_*$ is evaluated at the present epoch $z=0$.  These formula
also provide an accurate inverse relation $M_h \rightarrow M_v$.  Note 
that in the limit $c(M_v) \rightarrow \infty$, the mass correction is
small and in the limit $c(M_v) \rightarrow 0$, it is independent of
$c$.  Therefore, to obtain an accurate inverse mapping
one can utilize $c(M_h)$ in the inversion equation 
(\ref{eqn:massmap}) to obtain $M_v(M_h)$ to $\sim 1\%$ across all 
concentrations.  For better accuracy one can iterate the procedure
as $c(M_v(M_h))$.

\end{document}